\documentclass[journal]{IEEEtran}

\ifCLASSINFOpdf
\else
   \usepackage[dvips]{graphicx}
\fi
\usepackage{url}
\usepackage{bbm}
\usepackage{adjustbox}
\usepackage{balance}
\usepackage{glossaries}
\usepackage{graphicx}
\usepackage{cite}
\usepackage{bm}
\usepackage{mathtools}
\usepackage{amsmath}
\newacronym{MAP}{MAP}{Maximum a Posterior}
\newacronym{LM}{LM}{Lagrange multiplier}
\newacronym{ML}{ML}{Maximum Likelihood}
\newacronym{JMAP-ML}{JMAP-ML}{Joint Maximum A Posterior-Maximum Likelihood}
\newacronym{EKF-SA}{EKF-SA}{EKF-based State Augmentation}
\newacronym{EKS-SA}{EKS-SA}{EKF-based State Augmentation}
\newacronym{EM}{EM}{Expectation Maximization}
\newacronym{CD}{CD}{Coordinate Descent}
\newacronym{CRLB}{CRLB}{Cram\'er Rao Lower Bound}
\newacronym{MSE}{MSE}{Mean Square Error}
\newacronym{i.i.d.}{i.i.d.}{independent and identically distributed}
\newacronym{FI}{FI}{Fisher's identity}
\newacronym{FIM}{FIM}{Fisher Information Matrix}
\newacronym{MC}{MC}{Monte Carlo}
\newacronym{KF}{KF}{Kalman filter}
\newacronym{EKF}{EKF}{extended Kalman filter}
\newacronym{UKF}{UKF}{unscented Kalman filter}
\newacronym{PF}{PF}{particle filter}
\newacronym{KS}{KS}{Kalman smoother}
\newacronym{EKS}{EKS}{extended Kalman smoother}
\newacronym{RTS}{RTS}{Rauch–Tung–Striebel}
\newacronym{GD}{GD}{gradient descent}
\newacronym{BCLS}{BCLS}{Bias Compensated Least Squares}
\newacronym{BCKS}{BCKS}{Bias Compensated Kalman Smoother}
\newacronym{WTLS}{WTLS}{weighted total Least Squares}
\newacronym{EIV}{EIV}{Errors-in-Variables}
\newacronym{PEIV}{PEIV}{Partial Errors-in-Variables}
\newacronym{TLS}{TLS}{total Least Squares}
\newacronym{RMSE}{RMSE}{root mean square error}
\newacronym{PEM}{PEM}{Prediction Error Minimization}
\newacronym{LS}{LS}{Least Squares}
\makeglossaries

\newcommand{\txo}{{\text{o}}}

\begin{document}

\title{Joint State and Parameter Estimation Using the Partial Errors-in-Variables Principle}

\author{Peng Liu, Kailai Li,  Gustaf Hendeby,  and Fredrik Gustafsson
\thanks{The work is funded in part by the Swedish Research Council under the grant Scalable Kalman Filters and in part by ZENITH of Link\"oping University under the grant Computational Agile Sensing and Inference for Intelligent Systems.}
\thanks{Peng Liu, Gustaf Hendeby, and Fredrik Gustafsson are with the Division of Automatic Control, Department of Electrical Engineering, Link\"oping University, Link\"oping, Sweden. Email: {\tt\small {\{peng.liu, gustaf.hendeby, fredrik.gustafsson\}}@liu.se}}
\thanks{Kailai Li is with the Bernoulli Institute for Mathematics, Computer Science and Artificial Intelligence, University of Groningen, Groningen, the Netherlands. Email: {\tt\small{kailai.li@rug.nl}}}
}

\markboth{Journal of \LaTeX\ Class Files, Vol. xx, No. x, xx}
{Shell \MakeLowercase{\textit{et al.}}: Bare Demo of IEEEtran.cls for IEEE Journals}
\maketitle

\begin{abstract}
This letter proposes a new method for joint state and parameter estimation in uncertain dynamical systems. We exploit the partial errors-in-variables (PEIV) principle and formulate a regression problem in the sense of weighted total least squares, where the uncertainty in the parameter prior is explicitly considered. Based thereon, the PEIV regression can be solved iteratively through the Kalman smoothing and the regularized least squares for estimating the state and the parameter, respectively. The simulations demonstrate improved accuracy of the proposed method compared to existing approaches, including the joint maximum a posterior-maximum likelihood, the expectation maximisation, and the augmented state extended Kalman smoother.
\end{abstract}


\IEEEpeerreviewmaketitle

\section{Introduction}
Estimating states of uncertain dynamical systems plays fundamental roles in statistical signal processing and has various application scenarios, such as localization, tracking, energy, and robotics~\cite{gustafsson2002particle, roth2014ekf, ghahremani2011dynamic, simanek2014evaluation}, etc. Conventionally, state estimation problems can be solved recursively, either online using the Kalman filter and its derivatives, such as the \gls{EKF}~\cite{anderson2012optimal}, or offline based on the smoothing techniques, such as the extended Kalman smoother (EKS), for enhanced estimation accuracy~\cite{rauch1965maximum,sarkka2023bayesian}. 

However, standard filtering and smoothing algorithms assume the complete knowledge of the models, which is hard to reach in practice. A more realistic but more challenging scenario involves state-space modeling with unknown or uncertain parameters. One strategy to mitigate this issue is to augment the state with the parameter for joint estimation within the framework of EKF or EKS~\cite{ljung1979asymptotic}. While the resulting augmented state EKF or EKS have become popular owing to its simplicity, they may suffer from poor estimation accuracy due to observability degradation~\cite{yu2016improved}. Alternatively, iterative estimation methods have been investigated for joint state and parameter estimation, such as the maximum likelihood (ML) method~\cite{kay1993fundamentals}, which demonstrates favorable asymptotic properties and has been applied for state-space models together with the expectation maximisation (EM) algorithm~\cite{dempster1977maximum}. However, the ML method may deliver biased parameter estimates and fail to reach the Cramér–Rao bound given small datasets~\cite{yeredor2000joint}. This issue can be mitigated by the joint maximum a posterior-maximum likelihood (JMAP-ML) method involving numerical optimisation, such as the coordinate descent algorithm~\cite{wright2015coordinate}. This method has been widely exploited in many tasks including sensor calibration, epidemic modeling, and robust localization~\cite{kok2014maximum,liu2022joint,yin2013and}. However, the JMAP-ML disregards the uncertainty in the parameter prior, which may lead to insufficient accuracy~\cite{yeredor2000joint}.

In this paper, we investigate the possibility of explicitly incorporating the uncertainty in the parameter prior for joint state and parameter estimation of linear models, where the partial errors-in-variables (PEIV) model is exploited for regression. The standard errors-in-variables (EIV) model contains a regressor matrix that is subject to noise corruption \cite{fuller2009measurement,liu2023weighted}, which can be handled by the total least squares (TLS) for i.i.d. regressor and measurement noises~\cite{golub1980analysis} or the weighted total least squares (WTLS) for more general noise patterns~\cite{amiri2012weighted,schaffrin2008weighted}. For the PEIV model, where the regressor matrix is partially uncertain, it is possible to reformulate the model w.r.t. the uncertain part and apply WTLS for regression~\cite{xu2012total}. To the best of the authors' knowledge, there is no existing literature that investigates joint state and parameter estimation problem from an errors-in-variables perspective.

\subsection*{Contribution}
We propose a novel iterative framework for joint state and parameter estimation based on the partial errors-in-variables (PEIV) principle, which explicitly addresses the uncertainty in parameter prior. The joint estimation problem is formulated in the sense of WTLS and solved iteratively through the Kalman smoothing and the regularized least squares for estimating the state and parameter, respectively. The proposed method is evaluated through Monte Carlo simulation. Numerical results show its improved parameter estimation accuracy in comparison with the JMAP-ML, the EM, and the augmented state extended Kalman smoother (ASEKS).

The remainder of the paper is organized as follows. Sec.~\ref{PF} provides the signal model, followed by an overview of existing methods in Sec.~\ref{Previous Method} and \ref{joint method}. Sec.~\ref{KS PEIV} introduces the proposed PEIV-based framework, and Sec.~\ref{simulation} presents the numerical simulation. Finally, conclusions are drawn in Sec.~\ref{conclusion}.

\section{Signal Model}\label{PF}
To make the derivations explicit, we will assume a state-space model that is linear  in both the state and parameters 
\begin{equation}
    \begin{aligned}
        x_{k+1} &= F(\theta^\txo) x_{k} + v_{k}\,,\\
        y_k &= H(\theta^\txo)x_k + e_k\,,
    \end{aligned}
    \label{linear example}
\end{equation}
where the state-space matrices $F(\theta^{\txo})$ and $H(\theta^{\txo})$ are linear functions of the true parameter value $\theta^{\txo} \in \mathbbm{R}^d$ given as
\begin{equation}
\begin{aligned}
    F(\theta^{\txo}) &= F_0+\sum_{i=1}^d \theta^{\txo}_i F_i\quad \text{and} \\
    H(\theta^{\txo}) &= H_0+\sum_{i=1}^d \theta^{\txo}_i H_i\,,
\end{aligned}
    \label{linear parameter model}
\end{equation}
respectively. The matrices $F_i$ and $H_i$ are assumed to be known. $x_k\in\mathbbm{R}^n$ denotes the state vector, and $y_k\in\mathbbm{R}^{m}$ is the measurement. $v_{k}$ and $e_k$ are the white Gaussian-distributed process and measurement noises of covariance matrices $Q$ and $R$, respectively.   $\theta^{\txo}_i$ denotes the $i$-th element in the parameter vector. Further, $x_k$, $v_{k}$, and $e_k$ are assumed to be mutually independent. The initial state and the parameter priors are assumed to be Gaussian-distributed with
\begin{equation}
\begin{aligned}
    x^{\txo}_0 &\sim\mathcal{N}(m_0,P_0)\quad \text{and} \\
    \hat{\theta} &\sim \mathcal{N}(\theta^{\text{o}},\Sigma_\theta)\,,
\end{aligned}
\label{distributions}
\end{equation}
respectively. Let $X^\text{o}=[\,(x^{\text{o}}_0)^\top,\dots,(x^{\text{o}}_N)^\top\,]^\top$ contain all the state vectors and $Y=[\,y_1^\top,\dots,y_N^\top\,]^\top$ all the measurement. $x^{\txo}_k$ denotes the true state. Using a prior on the state and parameter allows the MAP approach maximising $P(X,\theta|Y)$, but we compare to the EM approach that maximises $P(Y|\theta)$ and the JMAP-ML method that maximises  $P(Y,X|\theta)$ (MAP and ML for estimating $X$ and $\theta$, respectively).
 
\section{Separate state and parameter estimation}\label{Previous Method}
The proposed PEIV method as well as the EM and JMAP-ML method all lead to algorithms that iteratively estimate the state and parameter. For that purpose, we derive the fundamental estimation modules in this section. These are rather straightforward to derive, and the main issue is to re-formulate the model given by \eqref{linear example} and \eqref{linear parameter model} to the following linear regression models
\begin{equation}
\begin{aligned}
    \bar{Y}&=\Psi(\theta^{\txo})X^{\txo}+\eta\,\quad\text{or} \\
    \quad\bar{Y}&=\Phi(X^{\txo}) \theta^{\txo} + c(X^{\txo}) + \eta\,.
\end{aligned}
    \label{parameter with error}
\end{equation}
Here, $c(X^\txo)$ represents the component that is independent of $\theta^{\txo}$. The interpretation of it will be provided later. The first model formulation leads to the Kalman smoother for a given parameter, and the second one induces the least squares estimate of the parameters $\theta^{\txo}$, given the state sequence $X^{\txo}$.

\subsection{Kalman Smoother}\label{ordinary KS}
The Kalman smoother (KS) can be formulated as a MAP problem given by
\begin{equation}
    \begin{aligned}
        \hat{X} =& \arg \max_X \log P(X|Y)\\
        =& \arg \max_X \sum_{i=1}^{N}\log P(y_i|x_i) +  \sum_{j=1}^{N}\log P(x_j|x_{j-1}) \\
        &+ \log P(x_0)\,.
    \end{aligned}
    \label{MAP smoothing}
\end{equation}
By exploiting the model (\ref{linear example}), (\ref{MAP smoothing}) can be expressed as
\begin{equation}
    \begin{aligned}       
       \hat{x}_{0:N} =\arg \min_X \Big\{& \|Y-C(\theta^{\txo})X\|^2_{\mathbf{R}^{-1}} \\
        +&\|A(\theta^{\txo})X\|^2_{\mathbf{Q}^{-1}}+ \|x_0-m_0\|^2_{P^{-1}_0}\Big\}\,, 
    \end{aligned}
    \label{KS MAP formulation}
\end{equation}
where $m_0$ and $P_0$ denote the mean and covariance of the initial state prior $x_0$, respectively. To achieve a conciser formulation, we introduce $\mathbf{R}=\mathbf{diag}(R,\dots,R)$, $\mathbf{Q}=\mathbf{diag}(Q,\dots,Q)$, and $\|(\cdot)\|^2_W = (\cdot)^\top W (\cdot)$. $\mathbf{diag}$ denotes the diagonal matrix, and $A(\theta^{\txo})$ and $C(\theta^{\txo})$ are defined by
\begin{equation}
    \begin{aligned}
        A(\theta^{\txo}) &= \begin{bmatrix}
             F(\theta^{\txo}) &-\mathbf{I} & 0 &\dots&0 \\
            0  & F(\theta^{\txo}) & -\mathbf{I} & \dots & 0\\
            \dots & \dots & \dots & \dots & \dots\\
            0 & 0 & \dots & F(\theta^{\txo})&-\mathbf{I}  
        \end{bmatrix}\,,\\
        C(\theta^{\txo})&= 
        \begin{bmatrix}
            0 & H(\theta^{\txo}) & 0 &\dots &0 \\
            0 & 0 & H(\theta^{\txo}) & \dots & 0\\
            \dots & \dots & \dots & \dots & \dots\\
            0&0 & \dots & 0 & H(\theta^{\txo}) 
        \end{bmatrix}
        \,,
        \end{aligned}
        \label{A and Cd structure}
        \end{equation}
respectively. With these definitions,  (\ref{KS MAP formulation}) can be formulated as the solution to the following linear regression models
\begin{equation*}
    \begin{aligned}
        Y &= C(\theta^{\txo})X^{\txo} + E\,,\\
        0 &= A(\theta^{\txo})X^{\txo} + V\,,\\
        m_0 &= x^{\txo}_0 + \epsilon\,,
    \end{aligned}
\end{equation*}
where $\mathbf{cov}(E) = \mathbf{R}$, $\mathbf{cov}(V)=\mathbf{Q}$, and $\mathbf{cov}(\epsilon) = P_0$. $0$ denotes zero vector. These  equations can be summarized as follows
\begin{equation}
\begin{aligned}
        \bar{Y} &= \begin{bmatrix}
             Y  \\
             0 \\
             m_0 
        \end{bmatrix} =  \begin{bmatrix}
             C(\theta^\txo)  \\
             A(\theta^\txo) \\
             [\,\mathbf{I},\mathbf{0}\,]
        \end{bmatrix}X^{\txo} + \begin{bmatrix}
             E  \\
             V \\
             \epsilon
        \end{bmatrix} \\
        &= \Psi(\theta^{\txo})X^{\txo} + \eta\,.
\end{aligned}
    \label{KS compact linear form}
\end{equation}
Given the assumption of mutual independence for the initial state, and the process and measurement noises, we have $\mathbf{cov}(\eta)=\mathbf{cov}([\,E^\top,V^\top,\epsilon^\top]^\top) = \mathbf{diag}(\mathbf{R},\mathbf{Q},P_0)\eqqcolon\Sigma_{\eta}$. $\bar{Y}$ serves as an augmented measurement based on the prior of $x^{\txo}_0$. The state estimate can be determined by the least squares (LS) assuming that the parameter $\theta^{\txo}$ is known, namely,
\begin{equation}
    \begin{aligned}
        \hat{X} &= (\Psi(\theta^{\txo})^\top\Sigma^{-1}_\eta \Psi(\theta^{\txo}))^{-1} \Psi(\theta^{\txo})^\top \Sigma^{-1}_\eta \bar{Y}\,.
    \end{aligned}
    \label{KS estimates}
\end{equation}
The covariance matrix of the estimation error is given by 
\begin{equation}
    \Sigma_{X} = (\Psi(\theta^{\txo})^\top\Sigma^{-1}_\eta \Psi(\theta^{\txo}))^{-1}\,.
    \label{KS state covariance}
\end{equation}
For implementing the Kalman smoother in practice, the recursive forward-backward version is perferred and runs much faster than the batch-wise solution~\cite{rauch1965maximum}. We give the batch-wise formulation here for the sake of clearness, which also assists introducing the JMAP-ML method in Sec.~\ref{iterative method section}.

\subsection{Parameter Estimation}
\label{Regularized Least Squares}
To derive the parameter estimation solution, we first need to rewrite \eqref{KS compact linear form} as a linear regression in $\theta^{\txo}$, and not in $X^{\txo}$. 
It is straightforward to show that $\Psi(\theta^{\txo})X^{\txo}$ can be written as
\begin{equation}
\begin{aligned}
    \Psi(\theta^{\txo})X^{\txo}&=D(X^{\txo})\mathbf{vec}(\Psi(\theta^{\txo}))\,,\quad\text{with}\\
    D(X^{\txo})&=(X^{\txo})^\top\otimes\mathbf{I}\,.
\end{aligned}
    \label{kroneck and vectorize}
\end{equation}
$\otimes$ is the Kronecker product, and $\mathbf{vec}(\cdot)$ denotes the matrix vectorisation. (\ref{A and Cd structure}) and (\ref{KS compact linear form}) show that only a portion of the elements in $\Psi(\theta^{\txo})$ is a function of $\theta^{\txo}$, whereas the others are independent of $\theta^{\txo}$. Accordingly, $\mathbf{vec}(\Psi(\theta^{\txo}))$ can be reformulated into 
\begin{equation}
    \begin{aligned}
        \mathbf{vec}(\Psi(\theta^{\txo}))&=h + B\theta^{\txo}\,.
    \end{aligned}
    \label{K separation}
\end{equation}
Combining \eqref{kroneck and vectorize} and \eqref{K separation} leads to
\begin{equation*}
\begin{aligned}
    \Psi(\theta^{\txo})X^{\txo} &= D(X^{\txo})\mathbf{vec}(\Psi(\theta^{\txo}))\\
    &=D(X^{\txo})h + D(X^{\txo})B\theta^{\txo}\\
    &=\Phi(X^{\txo})\theta^{\txo}+c(X^{\txo}).
\end{aligned}
\end{equation*}
Here, $c(X^{\txo}) = D(X^{\txo})h$. The solution to the parameter estimation problem in the sense of LS
\begin{equation}
    \hat{\theta}= \arg \min_{\theta} \|\bar{Y}-\Psi(\theta)X^{\txo}\|^2_{\Sigma^{-1}_\eta}\,
    \label{parameter in JMAP ML}
\end{equation}
can then be derived as
\begin{equation*}
    \begin{aligned}
        \hat{\theta} =&\,(B^\top D({X^{\txo}})^\top \Sigma^{-1}_\eta D({X}^{\txo})B)^{-1}\\
        &(B^\top D({X}^{\txo})^\top \Sigma^{-1}_\eta (\bar{Y}-D({X}^{\txo})h))\,,
    \end{aligned}
\end{equation*}
with covariance estimate
\begin{equation*}
\Sigma_{{\theta}} = (B^\top D({X}^{\txo})^\top \Sigma^{-1}_\eta D({X}^{\txo})B)^{-1}\,.
\end{equation*}

\section{Joint State and Parameter Estimation}\label{joint method}
State and parameter estimation can be iterated in different ways. This section provides overviews to well-known methods, before we introduce the PEIV method in the next section.

\subsection{Joint Maximum A Posterior-Maximum Likelihood}
\label{iterative method section}

 In this subsection, we explore the JMAP-ML method for estimating both the state and the model parameter iteratively. It aims to solve the optimisation problem given by
\begin{equation}
    \begin{aligned}
        \{\hat{X},\hat{\theta}\} =& \arg\min_{X,\theta}\,\, \log P(Y,X|\theta)\\
        =&\arg \min_{\theta}\{\arg\min_{X}\{\log P(X|Y,\theta)\}+\log P(Y|\theta)\}\,,
    \end{aligned}
    \label{JMAP ML cost function}
\end{equation}
where the parameter $\theta$ is a deterministic parameter, and the state $X$ is a random vector. This problem can be iteratively computed with an initialisation $\hat{\theta}^{1}$ following~\cite{yeredor2000joint}
\begin{equation}
    \begin{aligned}
        \hat{X}^{i+1} &= (\Psi(\hat{\theta}^i)^\top\Sigma^{-1}_\eta \Psi(\hat{\theta}^i))^{-1} \Psi(\hat{\theta}^i)^\top \Sigma^{-1}_\eta \bar{Y}\,,\\
        \hat{\theta}^{i+1} &= (B^\top D(\hat{X}^{i+1})^\top \Sigma^{-1}_\eta D(\hat{X}^{i+1})B)^{-1}\\
        &(B^\top D(\hat{X}^{i+1})^\top \Sigma^{-1}_\eta (\bar{Y}-D(\hat{X}^{i+1})h))\,,
    \end{aligned}
    \label{JMAP-ML iterative formula}
\end{equation}
where the KS and the LS are exploited for updating the state and parameter, respectively.

\subsection{Expectation Maximisation}
The JMAP-ML method discussed in Sec.~\ref{iterative method section} only utilizes the mean of the state estimate, and the covariance estimate is ignored. To fully exploit the information from state estimation, the EM method can be deployed.
It optimises for the parameter and state iteratively in the following ML problem 
\begin{equation}
    \begin{aligned}
        \hat{\theta} &= \arg \max_\theta \log P(Y|\theta)\,.
    \end{aligned}
    \label{ML cost function}
\end{equation}
The absence of state $X$ makes (\ref{ML cost function}) difficult to solve directly. The EM algorithm tackles this in two steps, namely, the $E$ step and $M$ step. The $E$ step estimates the state according to
\begin{equation}
    \mathcal{Q}(\theta,\hat{\theta}^i) = \mathbf{E}_{P(X|Y,\hat{\theta}^i)}(\log P(Y,X|\theta))\,,
    \label{E step}
\end{equation}
where $\hat{\theta}^{i}$ denotes the parameter estimate in the $i$-th iteration. The posterior distribution $P(X|Y,\hat{\theta}^{i})$ can be solved using KS introduced in Sec. \ref{ordinary KS}, with $\theta^{\txo}$ substituted by its estimate $\hat{\theta}^{i}$. After the $E$ step, the $M$ step updates $\hat{\theta}^i$ following 
\begin{equation}
    \hat{\theta}^{i+1}=\arg \max_\theta \mathcal{Q}(\theta,\hat{\theta}^i)\,.
    \label{M step}
\end{equation}
After updating the parameter in \eqref{M step}, we go back to the $E$ step in \eqref{E step} and repeat until convergence. In summary, the method resembles the one in \eqref{JMAP-ML iterative formula}. The only difference lies in iterating the parameter, where the uncertainty in the state estimate is considered as follows
\begin{equation}
    \begin{aligned}
    \hat{\theta}^{i+1} = \,&(\mathbf{E}(B^\top D(\hat{X}^{i+1})^\top \Sigma^{-1}_\eta D(\hat{X}^{i+1})B))^{-1}\\
    &\mathbf{E}(B^\top D(\hat{X}^{i+1})^\top \Sigma^{-1}_\eta (\bar{Y}-D(\hat{X}^{i+1})h))\,.
    \end{aligned}
    \label{EM algorithm}
    \end{equation}
Here, the expectation is computed with respect to $P(X|Y,\hat{\theta}^i)$.

\section{PEIV-based State and Parameter Estimation}
\label{KS PEIV}
We now introduce how to exploit the partial errors-in-variables (PEIV) modeling to facilitate joint state and parameter estimation. The regression on states in \eqref{KS compact linear form} contains a partially unknown regressor matrix $\Psi(\theta^{\txo})$ due to the uncertainty when estimating parameter $\theta^{\txo}$ in \eqref{A and Cd structure}. Based on \eqref{distributions} and \eqref{KS compact linear form}, we formulate the following WTLS problem to jointly estimate the state $X^{\txo}$ and the parameter $\theta^{\txo}$
\begin{equation*}
    \begin{aligned}
    \{\hat{\theta},\hat{X}\}&=\arg \min_{\theta,\eta}\bigg\Vert\begin{bmatrix}
         \theta-\hat{\theta}^1 \\
         \eta
    \end{bmatrix}\bigg\Vert^2_{\Sigma^{-1}},\,\\
    &s.t. \quad \eta = \bar{Y}-\Psi(\theta)X\,,
    \end{aligned}
\end{equation*}
where $\Sigma = \mathbf{diag}(\Sigma_\theta,\Sigma_\eta)$. It is straightforward to reformulate the objective by replacing $\eta$ with the constraint. This leads to
\begin{equation}
    J(\theta,X) =\|\theta-\hat{\theta}^1\|^2_{\Sigma^{-1}_{\theta}}+\|\bar{Y}-\Psi(\theta)X\|^2_{\Sigma^{-1}_\eta}\,,
\label{cost function in PEIV}
\end{equation}
where the first term can be seen as a generalized Tikhonov regularizer, with $\hat{\theta}^1$ being the initialised parameter estimate~\cite{golub1999tikhonov}. At each iteration, we can update the state following the first equation in \eqref{JMAP-ML iterative formula}. Afterward, the parameter estimate can be updated via the regularized least squares given the iterated state estimate $\hat{X}$. For that, we derive the closed-form derivative of $J(\theta,\hat{X})$ w.r.t.\ $\theta$ and set it to $0$, yielding
\begin{equation}
    \begin{aligned}
        \hat{\theta} = N^{-1}&(\Sigma^{-1}_{\theta} \hat{\theta}^{1} + B^\top D(\hat{X})^\top \Sigma^{-1}_{\eta} (\bar{Y}-D(\hat{X})h))\,.
    \end{aligned}
    \label{theta estimation peiv}
\end{equation}
The notation $N$ in (\ref{theta estimation peiv}) follows
\begin{equation}
    N=\Sigma^{-1}_{\theta}+ B^\top D(\hat{X})^\top \Sigma^{-1}_{\eta} D(\hat{X})B\,.
\end{equation}
\eqref{JMAP-ML iterative formula} and \eqref{theta estimation peiv} should be implemented iteratively.
Once converged, we can compute the estimation covariance of state $X$ according to (\ref{KS state covariance}) with the parameter estimate $\hat{\theta}$. The estimation covariance of $\hat{\theta}$ can be obtained by reformulating (\ref{cost function in PEIV}) given the state estimate $\hat{X}$ as follows
\begin{equation}
    \begin{aligned}
        \begin{bmatrix}
             \hat{\theta}^1  \\
             \bar{Y}-D(\hat{X})h
        \end{bmatrix}
        &= \begin{bmatrix}
             \mathbf{I}  \\
             D(\hat{X})B 
        \end{bmatrix}\theta^{\txo}+\begin{bmatrix}
             \Tilde{\theta}^{1}\,  \\
             \eta \,
        \end{bmatrix}\,.
    \end{aligned}
    \label{PEIV equivalent model}
\end{equation}
Here, $\Tilde{\theta}^{1}$ is the initialisation error. This leads to the covariance matrix $\mathbf{cov}(\hat{\theta})= N^{-1}$.

\begin{figure}[t!] 
\centering 
\adjustbox{trim={0.07\width} {0.05\height} {0.06\width} {0.05\height},clip}{\includegraphics[width=0.58\textwidth]{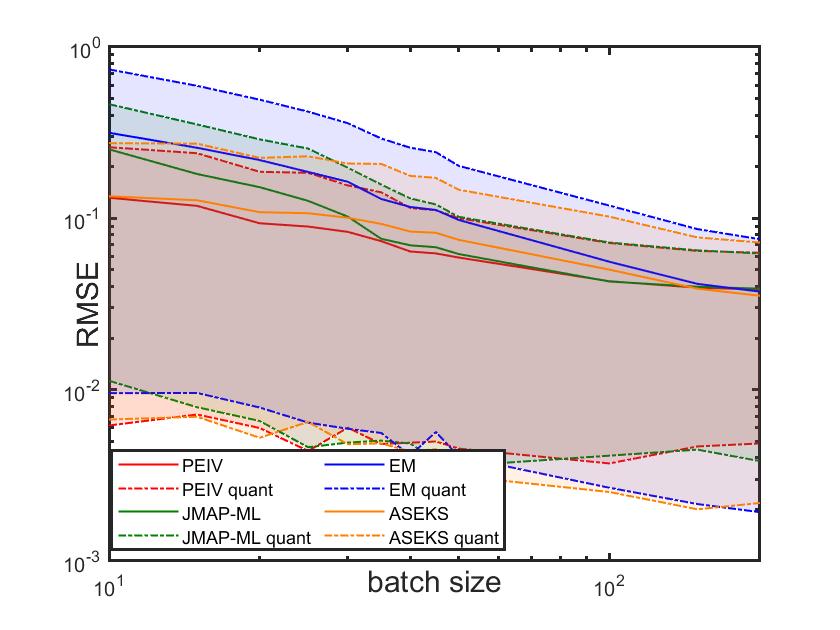}}
\caption{RMSE of parameter estimation w.r.t.\, batch size. The dashed lines in each color bounds the 5\% and 95\% quantiles given by each method.} 
\label{RMSE batch} 
\end{figure}

\section{Numerical Simulation}\label{simulation}
To demonstrate the merit of the PEIV principle in joint state and parameter estimation, we synthesize a numerical example with Monte Carlo simulation. We consider the following state-space model with scalar-valued state and parameter 
\begin{equation}
    \begin{aligned}
        x_{k+1} & =\theta^{\txo} x_k + v_k\,,\\
        y_k &= x_k + e_k\,.
    \end{aligned}
    \label{numerical example 1}
\end{equation}
The process noise $v_k$, the measurement noise $e_k$, and the initial state $x_0$ are assumed to be independent of each other, and we assume $v_k\sim\mathcal{N}(0,0.2)$ and $e_k\sim\mathcal{N}(0,0.09)$. We assume a stationary process with $x^{\txo}_k \sim \mathcal{N}(0,P)$, where 
\begin{equation*}
P=0.2/(1-(\theta^{\txo})^2)\,.
\end{equation*}
The state estimate is initialised as $\hat{x}_0\sim\mathcal{N}(y_1,2P)$, which implies that it is not necessary to know the prior. The true value of the parameter in the model is $\theta^{\txo}=0.9$. To quantify the estimation accuracy, we employ \gls{RMSE} criterion given by 
\begin{equation*}
\texttt{RMSE}_\theta =  \sqrt{\frac{1}{M}\sum_{i=1}^{M}(\hat{\theta}_i-\theta)^2}\,,
\end{equation*}
where $i$ denotes the $i$-th simulation (This equation shows the case for the parameter). We set the number of simulations $M=1000$. 

We evaluate our PEIV-based method with a focus on joint estimation using small batch size of data ranging within $\{10,15,20,25,30,35,40,45,50,100,150,200\}$ time steps. Three other state-of-the-art methods are involved for comparison, including the expectation maximisation (EM), the joint maximum a posterior-maximum likelihood (JMAP-ML), and the augmented state extended Kalman smoother (ASEKS) methods. 

\begin{figure}[t] 
\centering 
\adjustbox{trim={0.07\width} {0.05\height} {0.06\width} {0.05\height},clip}{\includegraphics[width=0.58\textwidth]{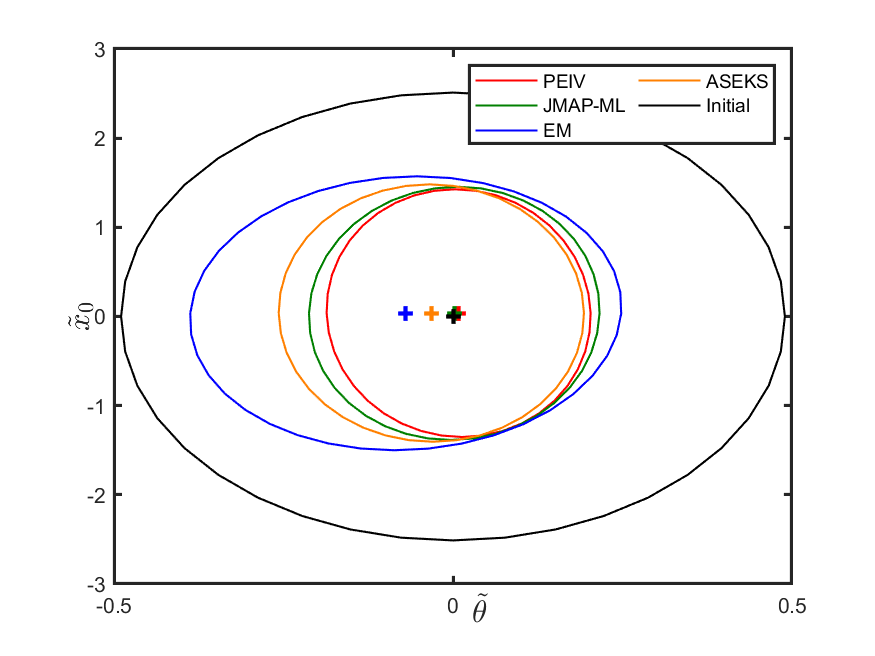}}
\caption{Error ellipse denoting the $95\%$ confidence interval for joint state and parameter estimation. Markers denote the biases of estimates.} 
\label{error ellipse}
\end{figure}

As shown in Fig.\,\ref{RMSE batch}, the proposed PEIV-based method outperforms all the other methods given small batch size of data ($\leq100$) in terms of RMSE and $95\%$ quantile. Additionally, we depict the error ellipses given by the simulations with a batch size of $30$ data points in Fig.\,\ref{error ellipse}, where $\tilde{x}_0$ and $\tilde{\theta}$ denotes the estimation errors of the initial state and the parameter, respectively. The proposed PEIV-based method delivers the best result in the benchmarking with the smallest error ellipse. 

\section{Conclusion}\label{conclusion}
In this letter, a novel principle for joint state and parameter estimation is proposed through the partial errors-in-variables modeling, where the uncertainty in the parameter prior is explicitly considered. Based thereon, we formulate the regression problem in the sense of WTLS, which is solved iteratively by the Kalman smoothing and the regularized least squares for updating the state and parameter, respectively. Numerical results based on simulations demonstrate that the proposed method outperforms state-of-the-art methods, including the EM, the JMAP-ML, and the ASEKS methods, in terms of estimation accuracy. 

For future investigation, we look forward to incorporating the uncertainty of state estimates into parameter estimation. Another possibility for extending the PEIV-based framework can be focused on tackling non-Gaussian noise patterns in state-space modeling.

\bibliographystyle{unsrt}
\balance
\bibliography{bibuseit.bib}

\end{document}